\documentclass[preprint,5p,times,twocolumn]{elsarticle}

\usepackage{amssymb}
\usepackage{amsmath}
\usepackage{xcolor}
\usepackage{enumitem}
\usepackage{fancyvrb}
\usepackage{threeparttable}
\usepackage{hyperref}
\usepackage{ulem}

\newcounter{bla}

\journal{Computer Physics Communications}

\begin{document}

\begin{frontmatter}



\title{Portal for High-Precision Atomic Data and Computation}


\author[b]{Amani Kiruga}
\author[a]{Charles Cheung\corref{cor1}}
\ead{ccheung@udel.edu}
\author[a]{Dmytro Filin}
\author[b]{Parinaz Barakhshan}
\author[e]{Akshay Bhosale}
\author[c]{Vipul Badhan}
\author[c,d]{Bindiya Arora}
\author[b]{Rudolf Eigenmann}
\author[a]{Marianna S. Safronova}

\cortext[cor1] {Corresponding author}
\address[a]{Department of Physics and Astronomy, University of Delaware, Delaware 19716, USA}
\address[b]{Department of Electrical and Computer Engineering, University of Delaware, Delaware 19716, USA}
\address[c]{Department of Physics, Guru Nanak Dev University, Amritsar, Punjab-143001, India}
\address[d]{Perimeter Institute for Theoretical Physics, Waterloo, Canada}
\address[e]{Department of Computer Science and Technology, University of Cambridge, Cambridge, UK}

\begin{abstract}
We've developed a scalable and sustainable online atomic data portal with an automated interface for easy update and addition of new data. The current portal provides energies, transition matrix elements, transition rates, radiative lifetimes, branching ratios, polarizabilities, hyperfine constants, and other data, for 28 atoms and ions. It also features an interactive polarizability plotting interface for neutral atoms and singly-charged ions. 
The data production is supported by recent developments of open-access atomic software based on our research codes, including new workflow algorithms, which allow large volumes of such data to be generated with automated accuracy assessments. 
This entails a new method of comparing our calculated values with data from the NIST Atomic Spectra Database.
All calculated values include estimated uncertainties. Data for more systems will be added in the future. Experimental values are included with references, where high-precision data are available. 
\\

%
%
%


\begin{keyword}
atomic structure \sep transition rates \sep polarizabilities

\end{keyword}

\end{abstract}

\end{frontmatter}


\section{Introduction}\label{sec:intro}
The goal of this project is to provide the scientific community with easily accessible high-quality atomic data and user-friendly, broadly-applicable modern relativistic atomic structure codes to treat electronic correlations. 
We have developed a free, open-access online portal (\href{https://www.udel.edu/atom}{https://www.udel.edu/atom}), which provides users with pre-computed atomic properties of various atoms and ions through an interactive interface, and enables them to easily view and download the data. This resource is hosted by the University of Delaware, and no registration is required to see the data.

The data are computed using the relativistic coupled-cluster (all-order) method  ~\cite{2007AO,2008Li,2008K,2011Ca,2011Rb,2013Be,2016Cs,2009Ra} and a combination of configuration interaction and coupled-cluster methods (CI+all-order)~\cite{PhysRevA.80.012516,2013Sr,2016Pb,2020Sr,PhysRevLett.133.023401,2019Cd,2021Sym,2025pCI}. These methods and corresponding codes are capable of calculating a very broad range of atomic properties to answer the significant needs of atomic, plasma, and astrophysics communities, among others. In order to facilitate the production of large volumes of atomic data for the portal, new workflow algorithms were developed to automate the atomic computations, analyze the resulting data, and process them for display on the portal. These new workflow algorithms are described in detail in~\ref{pCI-py}.
The CI+all-order codes have been published in Ref.~\cite{2025pCI} and are available on GitHub (\href{https://github.com/ud-pci/pCI}{https://github.com/ud-pci/pCI}).
Experimental values are displayed with references, where high-precision measurements data are available and expected to be more accurate than theory values. 

The portal provides the following atomic properties for various atoms and ions:
\begin{itemize}[noitemsep]
    \item energies
    \item transition matrix elements and rates
    \item radiative lifetimes
    \item branching ratios
    \item hyperfine constants
    \item quadrupole moments
    \item scalar and dynamic polarizabilities
\end{itemize}
A detailed description of these atomic data is given in Sec.~\ref{sec:theory_data}.

In the latest version released in April 2025 (Version 3.0 at the time of publication), we provide atomic properties of 28 atoms and ions. In addition, the data pipeline is automated. The pipeline begins with uploaded data, which are processed and used to update the database. It then automatically generates or updates the related web pages. This allows the portal to scale organically as new elements and data properties are added in upcoming updates. We also developed new workflow scripts to automatically generate and incorporate data for neutral Mg, Ca, and Sr, designed a new user interface, and added plots for polarizabilities, along with magic wavelengths \cite{2015Mg,2016Cs,2019Cd,2022Sr} and magic-zero (tune-out) wavelengths \cite{2012Rb,2021Cs}.

Version 1 of the portal was released in April 2021. In the first version of the portal, we included data for 12 monovalent atoms and ions: Li, Be$^+$, Na, Mg$^+$, K, Ca$^+$, Rb, Sr$^+$, Cs, Ba$^+$, Fr, Ra$^+$. In March 2022, we released Version 2, adding energies for the states of Li, Be$^+$, Mg$^+$, Ca$^+$, Rb, Sr$^+$, Ba$^+$, Fr, and Ra$^+$  not available in the \href{https://www.nist.gov/pml/atomic-spectra-database}{NIST Atomic Spectra Database}~\cite{NIST_ASD}. Data for 13 highly charged ions of interest for atomic clock development were also added. 

The portal access statistics are tracked via Google Analytics. Since the portal release, over 5,\,900 users from 95 countries have used the portal with 14,\,200 sessions and 88,\,500 pageviews.

\section{Theory and types of data}\label{sec:theory_data}

\begin{figure*}[ht!]
\includegraphics[width=\linewidth]{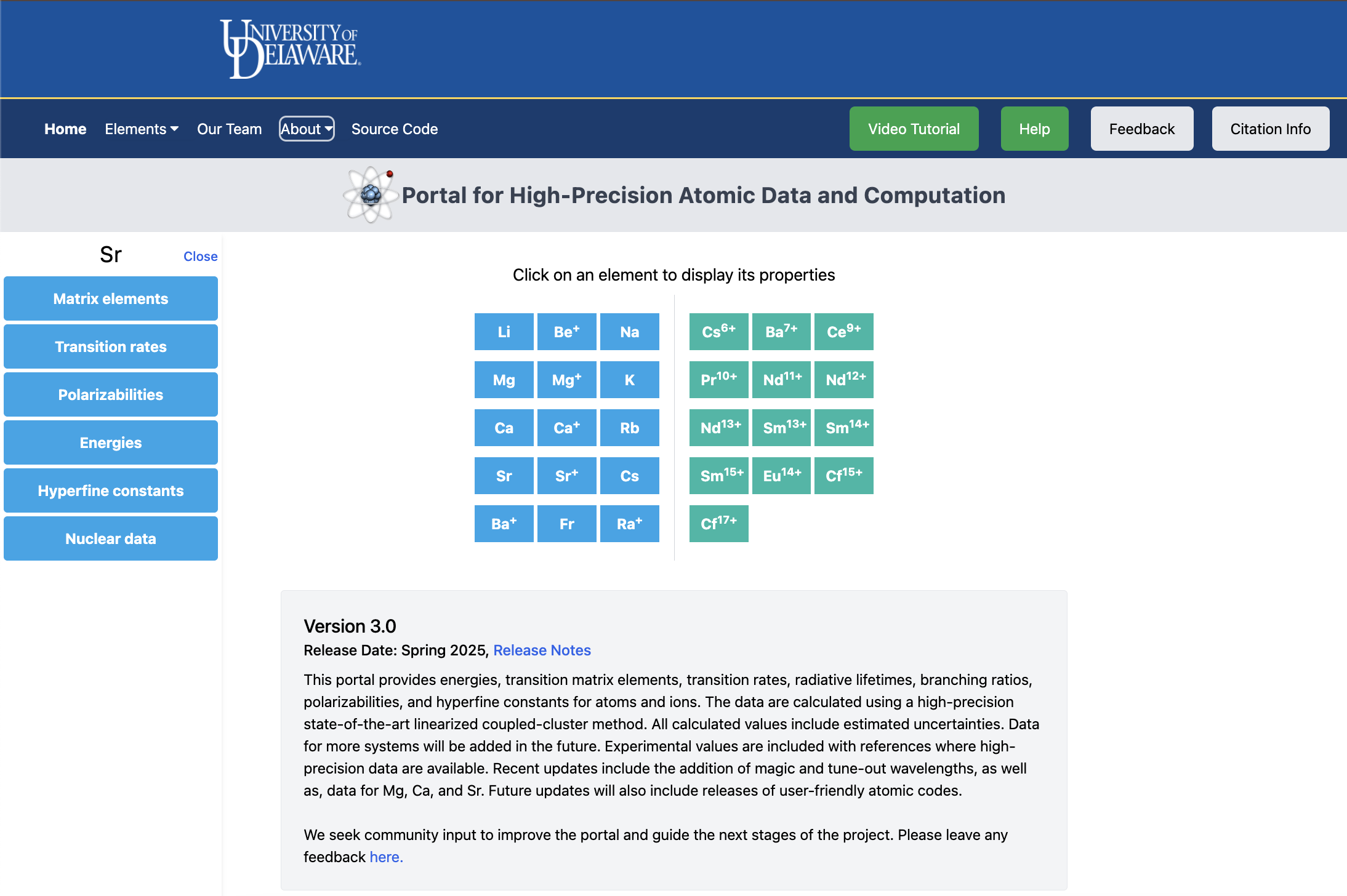}
\caption{\label{fig:homepage}Homepage of Version 3.0 of the portal. The user can select an atom or ion, which will display the available properties. URL: \href{https://www.udel.edu/atom}{https://www.udel.edu/atom}.}
\end{figure*}

\subsection{Energies}
Atomic energy levels are obtained by solving the many-electron Schr\"odinger equation
\begin{equation}
    H\Psi=E\Psi,
\end{equation}
where the wavefunction $\Psi$ is expanded depending on the approximation method used. We use the coupled-cluster (CC) and configuration interaction (CI) methods, which are described in Sec.~\ref{sec:theory_method}, to calculate energies for the portal. 
We provide recommended theoretical energies for states that are not available in the NIST database. These energies are in units of cm$^{-1}$ and include theoretical uncertainties. These recommended values, along with those from the NIST database, are used to compute the wavelengths required for calculations of transition rates, lifetimes and polarizabilities. For convenient access to the NIST database energies, we include direct links to the NIST Atomic Spectra Database~\cite{NIST_ASD} webpage for each element.

\subsection{Transition matrix elements and rates}

Matrix elements of irreducible tensor operators $T_q^k$ are evaluated using the Wigner-Eckart theorem \cite{johnson2007atomic}:
\begin{equation}
    \langle j^\prime m^\prime | T_q^k | j m\rangle = (-1)^{j^\prime-m^\prime}
    \left(
    \begin{array}{ccc}
        j^\prime & k & j \\ -m^\prime & q & m
    \end{array}
    \right)
    \langle j^\prime \Vert T^k \Vert j \rangle, 
\end{equation}
where $\langle j^\prime \Vert T^k \Vert j \rangle$ are reduced matrix elements independent of the magnetic quantum numbers $m$ and $m^\prime$, and projection $q$ of the tensor operator. 

Multipole transition probabilities $A_{wv}(Tk)$ from an upper state $v$ to a lower state $w$ are calculated from the wavelength and matrix elements using the following formulas:
\begin{equation}
    A_{wv}(E1)=\frac{2.02613\times10^{18}}{(2J_v+1)\lambda^3}S(E1),
\end{equation}
\begin{equation}
    A_{wv}(E2)=\frac{1.11995\times10^{18}}{(2J_v+1)\lambda^5}S(E2),
\end{equation}
\begin{equation}
    A_{wv}(E3)=\frac{3.14441\times10^{17}}{(2J_v+1)\lambda^7}S(E3),
\end{equation}
\begin{equation}
    A_{wv}(M1)=\frac{2.69735\times10^{13}}{(2J_v+1)\lambda^3}S(M1),
\end{equation}
\begin{equation}
    A_{wv}(M2)=\frac{1.49097\times10^{13}}{(2J_v+1)\lambda^5}S(M2),
\end{equation}
\begin{equation}
    A_{wv}(M3)=\frac{4.18610\times10^{12}}{(2J_v+1)\lambda^7}S(M3),
\end{equation}
where $J_v$ is the total angular momentum of the upper state $v$, $\lambda$ is the wavelength of the transition in \AA, and $S(Tk)=|\langle j_w \Vert T^k \Vert j_v \rangle|^2$ is the line strength of the transition. Here, $Tk$ is the electric or magnetic multipole operator, $Ek$ or $Mk$, where the ranks $k=1,2,3$ indicate dipole, quadrupole and octupole transitions, respectively. The portal displays reduced matrix elements in units of a.u. for electric multipole transitions, and in units of $\mu_0$ for magnetic multipole transitions. The uncertainties of the transition rate values are determined from the uncertainties in the matrix elements and wavelengths.

The radiative lifetime $\tau_v$ of an upper state $v$ is calculated as the inverse of the total transition probability
\begin{equation}
    \tau_v = \frac{1}{\sum_w A_{wv}},
\end{equation}
where the denominator contains the sum over all possible transition rates. The lifetimes are given in ns, unless specified. Note that the blackbody contribution is not included and may contribute outside of the quoted uncertainties for higher excited states. 
Branching ratio is calculated as the ratio of a transition rate to the total transition rate:
\begin{equation}
    R_{wv}=\frac{A_{wv}}{\sum_w A_{wv}}.
\end{equation}
This quantity is dimensionless.

Most of the transition probabilities provided by the portal are for electric-dipole transitions. Electric-quadrupole and magnetic-dipole transition rates, along with their corresponding matrix elements, are listed for transitions from metastable states. The buttons for the metastable states are of different color for easy identification. Other multipole transition probabilities are listed for highly-charged ions.

\subsection{Polarizabilities}
The total dynamic polarizability $\alpha(\omega)$  is computed as
\begin{equation}
    \alpha(\omega) = \alpha_0(\omega) + \chi(\theta)\alpha_2(\omega)\frac{3m^2-j(j+1)}{j(2j-1)},
\end{equation}
where $\theta$ is defined as the angle between the magnetic field $\mathbf{B}$ and the direction of light propagation determined by the wave vector $\mathbf{k}$ and $\chi(\theta)$ is given by
\begin{equation}
    \chi(\theta) = \frac{3\cos^2\theta - 1}{2}.
\end{equation}
The quantities $\alpha_0(\omega)$ and $\alpha_2(\omega)$ are the scalar and tensor parts of the polarizabilities computed for monovalent systems as
\begin{equation}
    \alpha_0(\omega)=\frac{2}{3(2j+1)}\sum\limits_{j^\prime}\frac{(E-E^\prime)|\langle j \Vert D \Vert j^\prime \rangle|^2}{\omega^2-(E-E^\prime)^2},
\end{equation}
and
\begin{multline}
    \alpha_2(\omega)=\sqrt{\frac{40j(2j-1)}{3(j+1)(2j+3)(2j+1)}} \\ \times\sum\limits_{j^\prime}(-1)^{j+j^\prime} 
    \left\{
    \begin{array}{ccc}
        j & 2 & j  \\
        1 & j^\prime & 1 
    \end{array}
    \right\} 
    \frac{(E-E^\prime)|\langle j \Vert D \Vert j^\prime \rangle|^2}{\omega^2-(E-E^\prime)^2},
\end{multline}
where $D$ is an electric dipole moment operator, $j^\prime$ represents the angular momentum of the intermediate states, and $E$ and $E^\prime$ are energies of the initial and intermediate states. 

For multivalent systems, we calculate the dynamic polarizabilities in the framework of CI+all-order. Here, the polarizability can be separated as $\alpha = \alpha^v + \alpha^c$,
where $\alpha^v$ and $\alpha^c$ are the valence and core contributions.

The expression for the valence polarizability $\alpha^v$ at a frequency $\omega$ for the state $|jm\rangle$ (where $j$ is the total angular momentum and $m$ is its projection) can be written (in a.u.) as
\begin{equation}
    \alpha^v(\omega)=2\sum_{j',m'} \frac{(E'-E)|\langle jm | D_z | j'm'\rangle|^2}{(E'-E)^2-\omega^2},
\end{equation}
where $D$ is an electric dipole moment operator and $E$ and $E'$ are the energies of the initial and intermediate states, respectively.
This expression can be rewritten as
\begin{eqnarray}
    \alpha^v(\omega) &=& \sum_{j',m'} \langle jm|D_z|j'm' \rangle\langle j'm' | D_z | jm \rangle  \nonumber\\
    &\times& \left[ \frac{1}{E'-E+\omega} + \frac{1}{E'-E-\omega} \right].\label{alpha1}
\end{eqnarray}

To find the valence contribution $\alpha^v$, we can rewrite this as 
\begin{equation}
    \alpha^v(\omega) =  \langle \Phi| D_z |\delta \phi_{+} \rangle + \langle \Phi| D_z |\delta \phi_{-} \rangle ,
\end{equation}
and use the Sternheimer~\cite{Ste50} or Dalgarno-Lewis~\cite{DalLew55} method and solve inhomogeneous equations
\begin{equation}
(H - E \pm \omega)\, |\delta \phi_{\pm} \rangle = d_z\, |\Phi \rangle,
\label{inhom}
\end{equation}
where $\Phi$ is an eigenstate of the Hamiltonian $H$. The wave function $|\delta \phi_{\pm} \rangle$ can be found from Eq.~\ref{inhom} as
\begin{eqnarray}
|\delta \phi_{\pm} \rangle &=& \frac{1}{H - E \pm \omega} D_z |\Phi \rangle \nonumber \\
&=& \sum_{j',m'} \frac{1}{H - E \pm \omega}|j'm'\rangle \langle j'm' |D_z |\Phi \rangle,
\label{phi}
\end{eqnarray}
where we used the closure relation $\sum_{j',m'} | j'm' \rangle \langle j'm' | = 1$.

The core contribution $\alpha^c$ can be calculated in the single-electron approximation using a sum-over-states approach. The single-electron matrix elements of the electric dipole operator include the random-phase-approximation (RPA) corrections. 

\begin{figure}
\includegraphics[width=\linewidth]{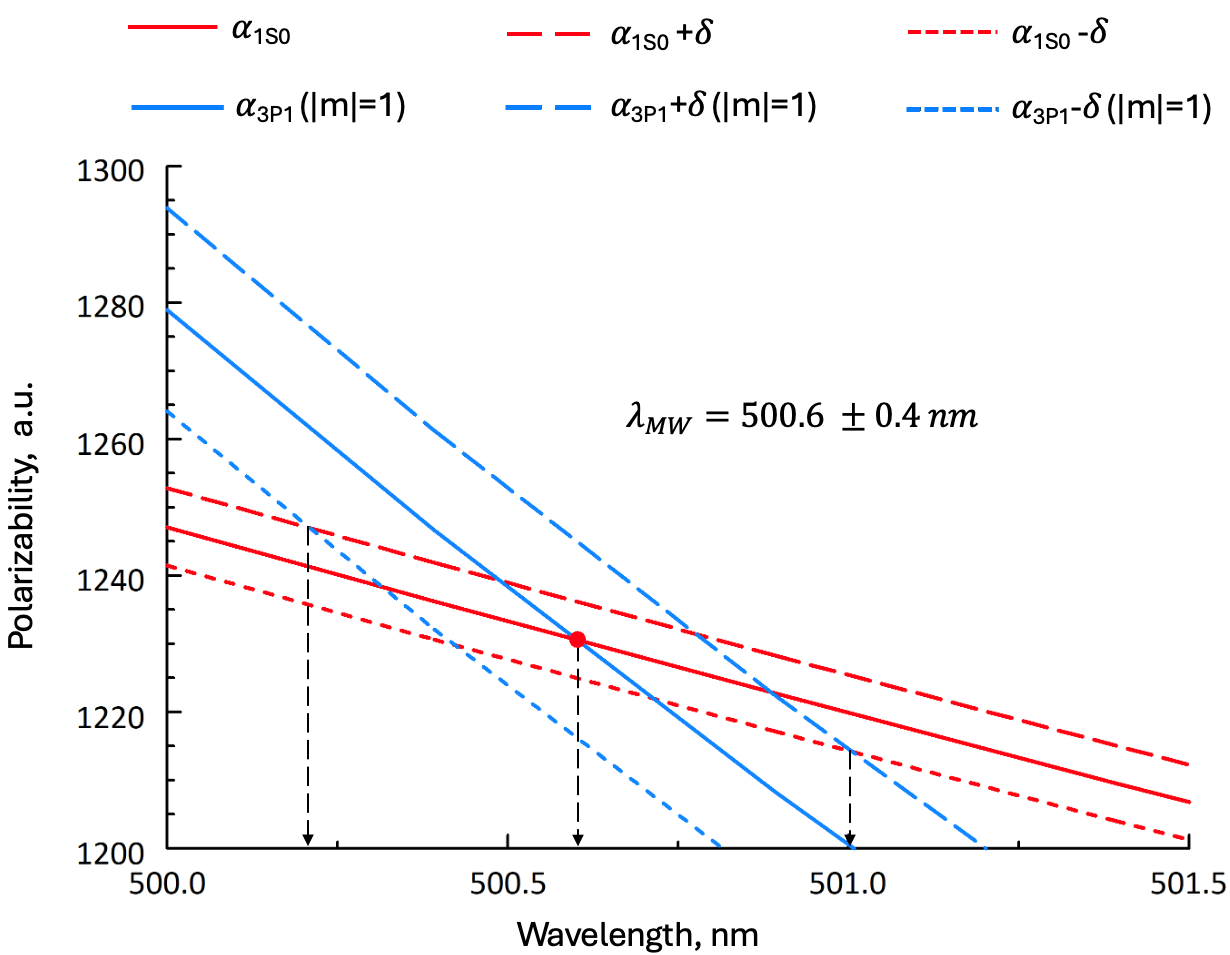}
\caption{Intersections of the polarizabilities $\alpha$, $\alpha-\delta$, and $\alpha+\delta$ computed for the $^1S_0$ and $^3P_1$ (with the magnetic quantum number $m=\pm1$) states of Sr in the vicinity of $\lambda=500.6$ nm.}
\label{magicwave}
\end{figure}

Based on the uncertainties of the matrix elements $\Delta$, we obtained the uncertainty $\delta$ of the atomic polarizability $\alpha$ as a function of the appropriate matrix elements.
To estimate the accuracy of magic wavelengths calculated for two atomic states, we implemented the following procedure. For each of these states, we build a graph for three polarizabilities as functions of a wavelength: $\alpha$, $\alpha-\delta$, and $\alpha+\delta$. Then, we approximate intersections of every polarizability curve with curves plotted for another state, as shown in Figure~\ref{magicwave}.  We consider the wavelengths for the most extreme intersections to be the bounds of the uncertainty of the magic wavelength $\lambda_{MW}$. As an example, in Figure~\ref{magicwave}, we presented the intersections in the vicinity of $\lambda=500.6$ nm computed for the $^1S_0$ and $^3P_1$ states of Sr (with the magnetic quantum number $m=\pm1$).

\subsection{Hyperfine constants}
The interaction of atomic electrons with the multipole moments of the nucleus
leads to a dependence of the atomic energy levels on a nuclear spin, referred to as the
atomic hyperfine structure \cite{johnson2007atomic}.
Our portal provides both theoretical and experimental values of the hyperfine constants $A$ (which quantifies interaction of atomic electrons with a nuclear magnetic dipole moment) in MHz for selected atoms, sourced from the published literature. 

In future updates, we plan to expand the datasets by incorporating additional theoretical values from new computations. Additionally, we aim to include hyperfine constants $B$, which quantifies interaction of atomic electrons with a nuclear electric quadrupole moment. This will offer a more comprehensive resource for researchers studying atomic structure, precision measurements, and quantum technologies.

\subsection{Nuclear data}
In addition to the detailed atomic properties, we also provide nuclear data  on the properties of isotopes with a half-life exceeding 10 minutes for various atoms. This dataset has been compiled from reliable and established nuclear data sources, including the works by Stone et al. \cite{stone,stone2}, Meija et al. \cite{meija}, and Angeli \cite{angeli}, among others. These references provide extensive nuclear data and are integral to our dataset, although additional citations will be added as needed. For the element Francium (Fr), due to its significance in parity-violating studies, we have made an exception and included isotopes with half-lives exceeding 3 minutes. 
The data provided includes the following properties:
\begin{itemize}[noitemsep]
    \item Nuclear spins
    \item Natural abundances
    \item Half-life for unstable isotopes
    \item Root-mean-square (rms) radii in fermi
    \item Recommended values of the magnetic dipole moment in $\mu_N$
    \item Recommended values of the electric quadrupole moment in $b$
\end{itemize}

\section{User interface}
\label{sec:UserInterface}

\begin{figure*}[ht!]
\includegraphics[width=\linewidth]{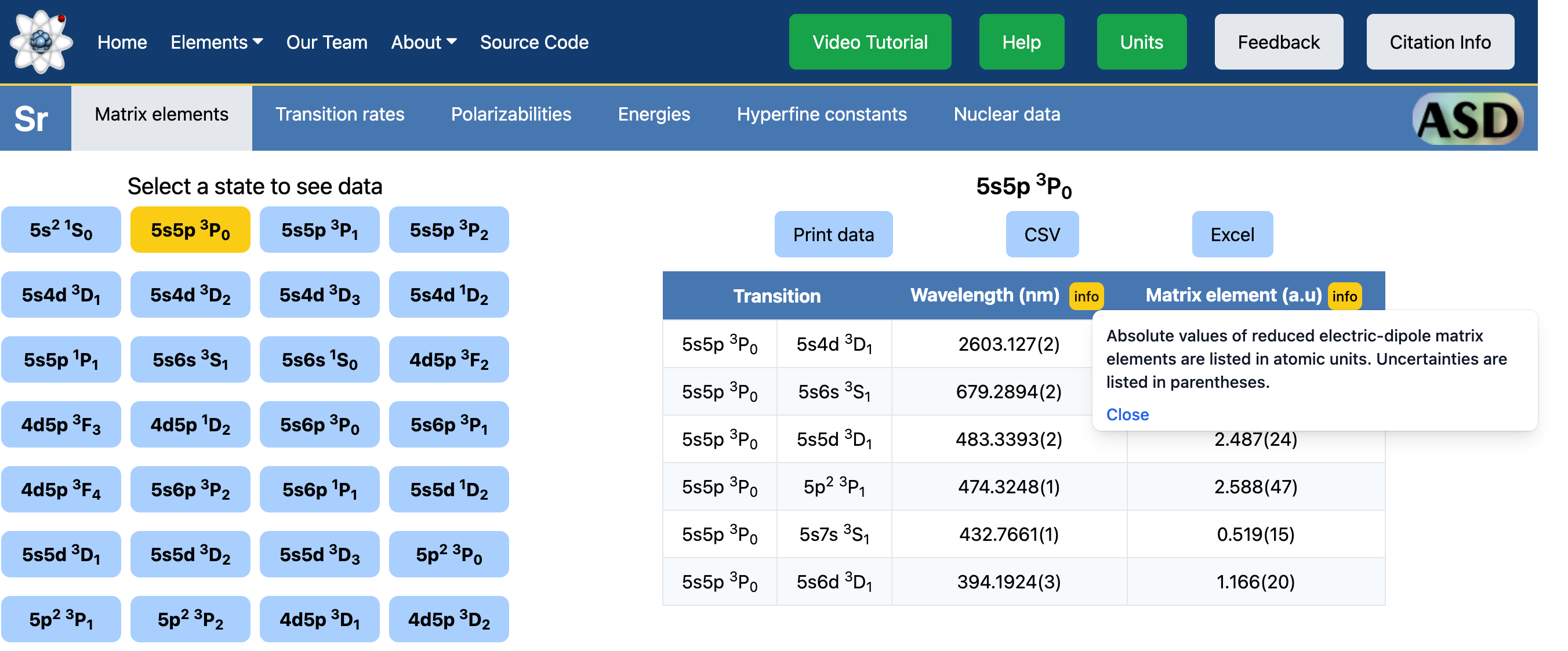}
\caption{\label{fig:matrixelements}A screenshot of the Matrix elements table for the Sr $5s5p\,^3\!P_0$ state showing a pop-up tooltip with information about the Matrix Elements column.}
\end{figure*}

\begin{figure*}[ht!]
\includegraphics[width=\linewidth]{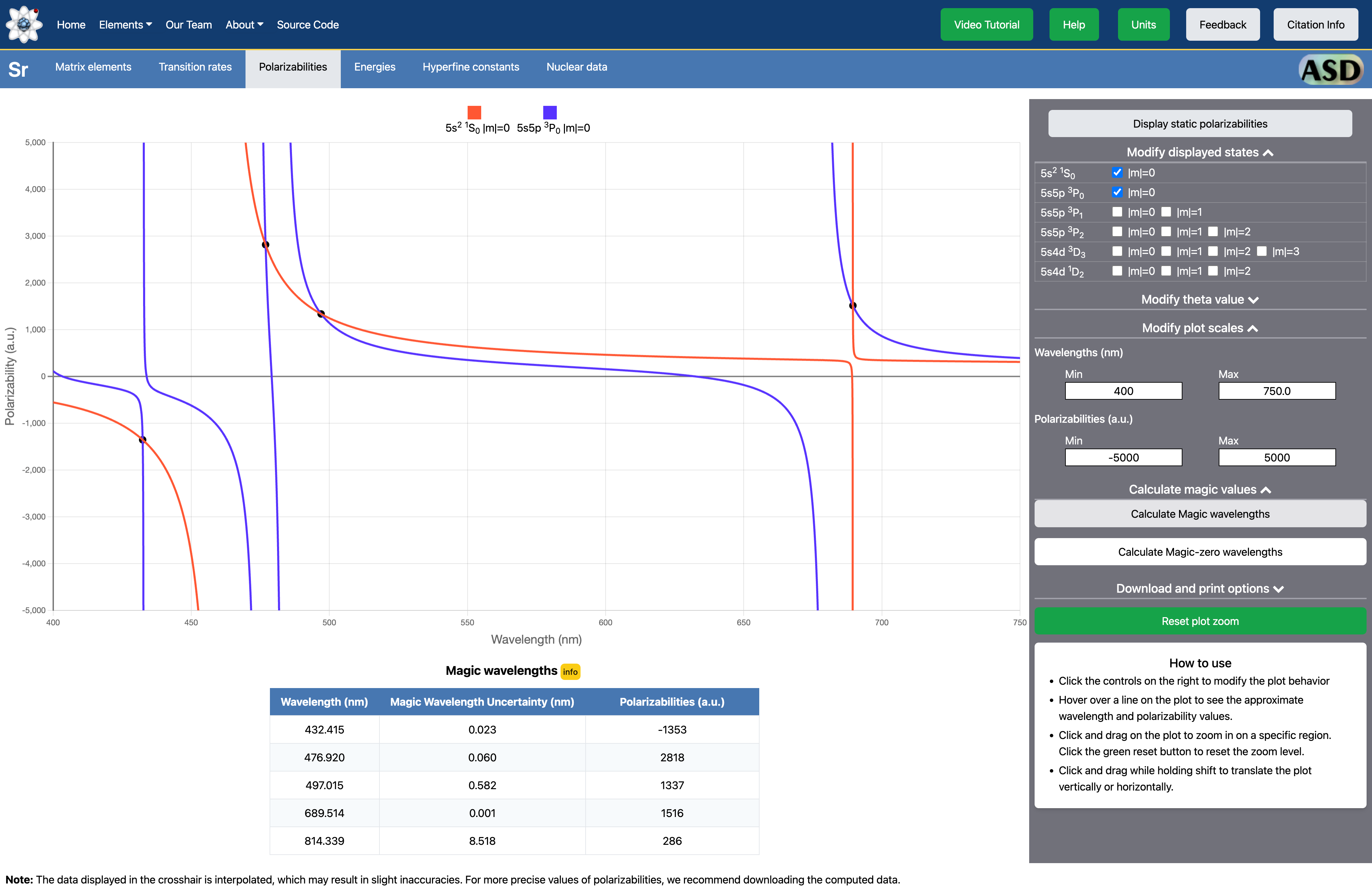}
\caption{A screenshot of the interactive polarizability plots for the Sr $5s^2\,^1\!S_0$ and $5s5p\,^3\!P_0$ states between 400 nm and 750 nm. Magic wavelengths (in nm), their uncertainties (in nm), and polarizabilities (in a.u.) are provided in a table below the plot. The panel on the right side allows users to modify the displayed states, $\theta$, and plot scales, as well as display the calculated magic wavelengths and magic-zero wavelengths. There is also a button to download the polarizability data. }
\label{fig:pol}
\end{figure*}

Figure~\ref{fig:homepage} shows the homepage of the portal. Here, the user is able to select an atom or ion, Sr in this case, which  displays the available properties. Most properties show a list of electron states; clicking on one typically shows the corresponding properties in tabular format. Figure~\ref{fig:matrixelements} shows the table of matrix elements for the Sr state $5s5p\,^3\!P_0$. Polarizability is an exception, which is displayed through interactive plots, as seen in Figure ~\ref{fig:pol}. 

Most tables are searchable, sortable, and filterable by atom, state, or data type. Data tables include buttons such as “info” and “Ref,” which provide inline explanations of the data and relevant sources, respectively. Pop-up tooltips appear next to most columns (see Figure~\ref{fig:matrixelements}, Column Matrix Element) offering details about the origin of the data and links to further information about how it was calculated. Each page features ``Help'' and ``Units'' buttons in the header that provide explanations about the page's functionality along with units and unit conversion parameters. A brief video walkthrough is provided to guide new users through the portal’s features. ``Print'' and ``Download'' functionality is also available on most tables.  Additionally, a feedback form is embedded directly into the interface, enabling users to submit suggestions or report issues to support continued development.

The portal interface is built with the React web design framework (nextjs.org) and features modular, responsive components for a streamlined user experience. Interactive visualizations are rendered using a custom modified version of the Chart.js library (chartjs.org), supporting zooming, panning, hover tooltips, and clickable legends. Mathematical expressions are displayed using KaTeX (katex.org) to ensure clear scientific notation. Magic and tune-out wavelengths are computed on demand via a Python Flask backend using the SciPy library, with results visualized in real time.

\section{Methods}\label{sec:theory_method}

\subsection{Coupled-cluster for monovalent systems}
Atomic properties of monovalent systems on the portal are calculated using a linearized coupled-cluster method~\cite{2007AO,2008Li,2008K,2011Ca,2011Rb,2013Be,2016Cs,2009Ra}. This method is also referred to as ``all-order'' in the literature, as it involves summing series of dominant many-body perturbation theory terms to all orders. In the single-double (SD) all-order approach, SD excitations of the Dirac-Fock orbitals are included. The single-double partial triples (SDpT) all-order approach includes classes of triple excitations. Omitted higher-order excitations are estimated by a scaling procedure which can be started from either SD or SDpT approximations. A detailed description of the SD and SDpT all-order approaches is given in Ref.~\cite{2007AO}. 

We carry out four all-order computations for electric-dipole matrix elements: (1) \textit{ab initio} SD, (2) SDpT, (3) scaled SD, and (4) scaled SDpT. Either SD or scaled SD data are taken as final values based on the comparison of different contributions to the matrix elements. An algorithm is used to determine the uncertainties of the electric-dipole matrix elements based on the spread of the four results, size of the correlation correction, and comparison of different contributions to the matrix elements. 

\subsection{CI methods for multivalent systems}
To calculate atomic properties of systems with a few valence electrons, we use the CI+all-order method that combines the configuration interaction and coupled-cluster methods ~\cite{PhysRevA.80.012516,2013Sr,2016Pb,2020Sr,PhysRevLett.133.023401,2019Cd,2021Sym,2025pCI}.  We include estimated uncertainties for energies and matrix elements, obtained by comparing the CI+all-order results with those from a corresponding CI+MBPT (many-body perturbation theory) method. For example, the values of divalent systems, such as neutral Sr, Ca and Mg, are obtained using the CI+all-order method, with uncertainties obtained as the difference between the CI+all-order and CI+MBPT results, with some additional uncertainty estimated due to the omission of additional small corrections~\cite{2013Sr}. More detail on the uncertainties is given below.

In general, we start by solving the Dirac-Hartree-Fock equations in the central-field approximation to construct one-electron basis functions for the core and valence electrons. The valence-valence correlation problem is solved using the CI method, while core correlations are included using the all-order approach. We use the pCI software package~\cite{2025pCI} for our computations. A detailed description of the code package, including details about the CI method, CI+MBPT and CI+all-order method, is given in Ref.~\cite{2025pCI} and references therein. 

Energies and matrix elements calculated via the CI+all-order approach include uncertainties, obtained by comparing results with those from the CI+MBPT approach. 
Due to the omission of additional small corrections, we set a minimum value on all uncertainties of computed matrix elements. The final uncertainty $\Delta$ for matrix elements is computed in quadrature using the formula:
\begin{equation}
\label{Delta}
    \Delta = \sqrt{\Delta_0^2+\Delta_\mathrm{min}^2},
\end{equation}
where $\Delta_0$ is the original uncertainty obtained as the difference between CI+all-order and CI+MBPT results, and $\Delta_\mathrm{min}$ is a minimum uncertainty value based on small corrections for $E1$ matrix elements beyond RPA~\cite{2013Sr}. 
For example, we set $\Delta^\mathrm{Sr}_\mathrm{min}=0.015$ for neutral Sr, $\Delta^\mathrm{Mg}_\mathrm{min}=0.003$ for neutral Mg, and $\Delta^\mathrm{Ca}_\mathrm{min}=0.013$ for neutral Ca. Energies are not affected by this minimum uncertainty value. 

\section{Data generation}\label{sec:data_gen}
Theoretical data displayed on the portal are generated by running the corresponding atomic computations with the pCI package~\cite{2025pCI}. In order to facilitate the production of large volumes of atomic data for the portal, we developed new workflow algorithms and scripts to automate the atomic computations. In addition to calculating data, these scripts also analyze the resulting data, correct any misidentifications, and process them for display on the portal. We leave the technical details of these algorithms and scripts to~\ref{pCI-py}. 

\subsection{Correspondence with NIST}\label{sec:correspondence}
\begin{figure*}[ht!]
\centering
\includegraphics[trim={0cm 0cm 0cm 5cm},clip,width=1\linewidth]{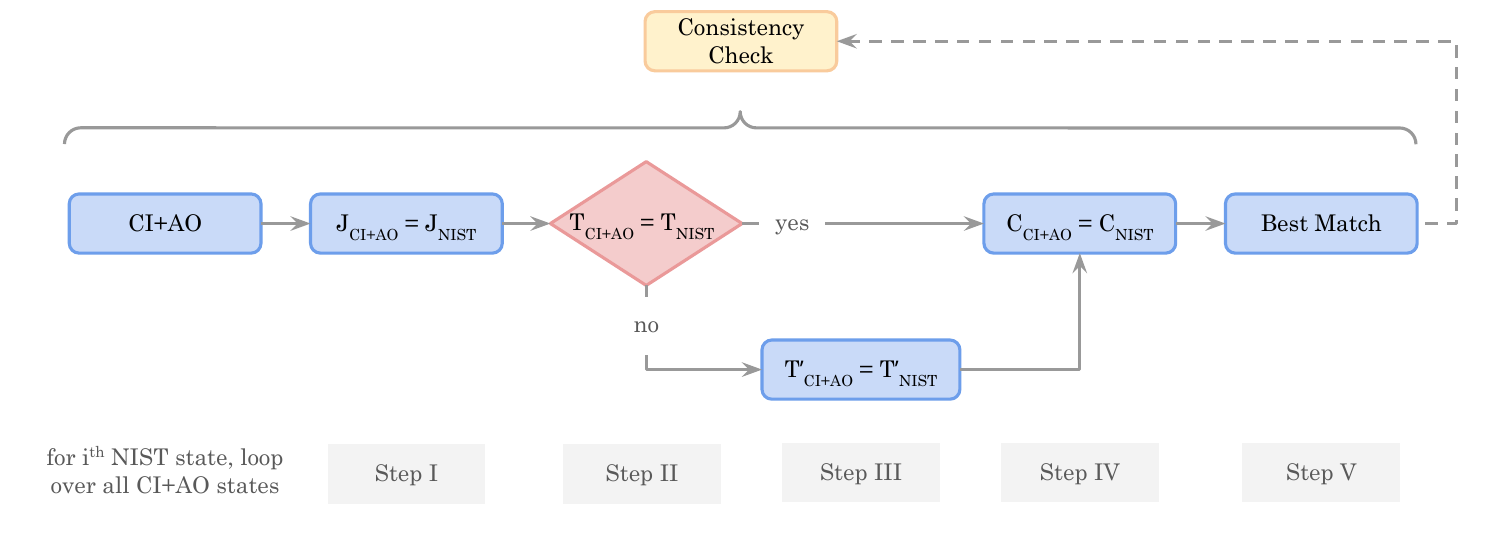}
\caption{The workflow for identifying the corresponding CI+all-order state for a given NIST state. For clarity, total angular momentum (J), term symbol (T), and electronic configuration (C) are denoted symbolically. ``CI+AO'' refers to the CI+all-order method. See Sec.~\ref{sec:correspondence} for details of the workflow.}
\label{correspondence_workflow}
\end{figure*}

Due to discrepancies in the basis set or numerical precision, the configurations and terms of energy levels outputted by the pCI computations may sometimes be misidentified. To account for this, we compare the computed data to high-precision experimental data, obtained by parsing the NIST Atomic Spectra Database~\cite{NIST_ASD} for the full list of energy levels. We compare the NIST and theoretical data computed with pCI, then establish a reliable correspondence, or mapping, between the states via a systematic five-step correspondence procedure, illustrated in Figure~\ref{correspondence_workflow}. This comparison also allows us to also provide data on the portal that are not available on the NIST database.

The correspondence procedure begins by analyzing the atomic states from the NIST database and pCI computations. For each ($i^{th}$) NIST state, we examine all states obtained from the pCI computations, and apply the following sequential filtering criteria:
\begin{enumerate}[noitemsep,label={\Roman*.}]
\item Identify states that share the same total angular momentum quantum number $J$ as the given NIST state.

\item From the subset obtained in Step I, select states that also match the atomic term symbol of the NIST entry. If found, proceed directly to step IV.

\item When there is no exact match for the term symbol, consider states where modifying the multiplicity by $\pm 1$ yields a match.

\item Further refine the selection by considering states in which either the primary or secondary electronic configuration aligns with that of the given NIST state.

\item The most appropriate match is selected from the candidates obtained in the previous step. Preference is given to states with the same primary electronic configuration. If multiple states share a primary configuration, the state with the smallest percentage difference in energy relative to the NIST value is chosen. If no primary configuration matches, the same criterion is applied to the secondary configuration.
\end{enumerate}

This procedure is applied iteratively to all NIST states. For each NIST state, the matched states are stored and used for additional \textit{consistency checks} in all stages of future iterations. These checks are made to avoid duplicate correspondences, ensure that no states are missed, maintain the correct ordering of principal quantum numbers in configurations for each term symbol, and other similar validations. This additional step ensures that the sequence of selected configurations remains physically meaningful and that no plausible states are overlooked. Any discrepancies identified during this step are manually corrected to maintain consistency. Additional states that are obtained by pCI that are not available at NIST are displayed on the portal with estimated theoretical uncertainties.

\begin{figure*}[ht!]
\includegraphics[width=\linewidth]{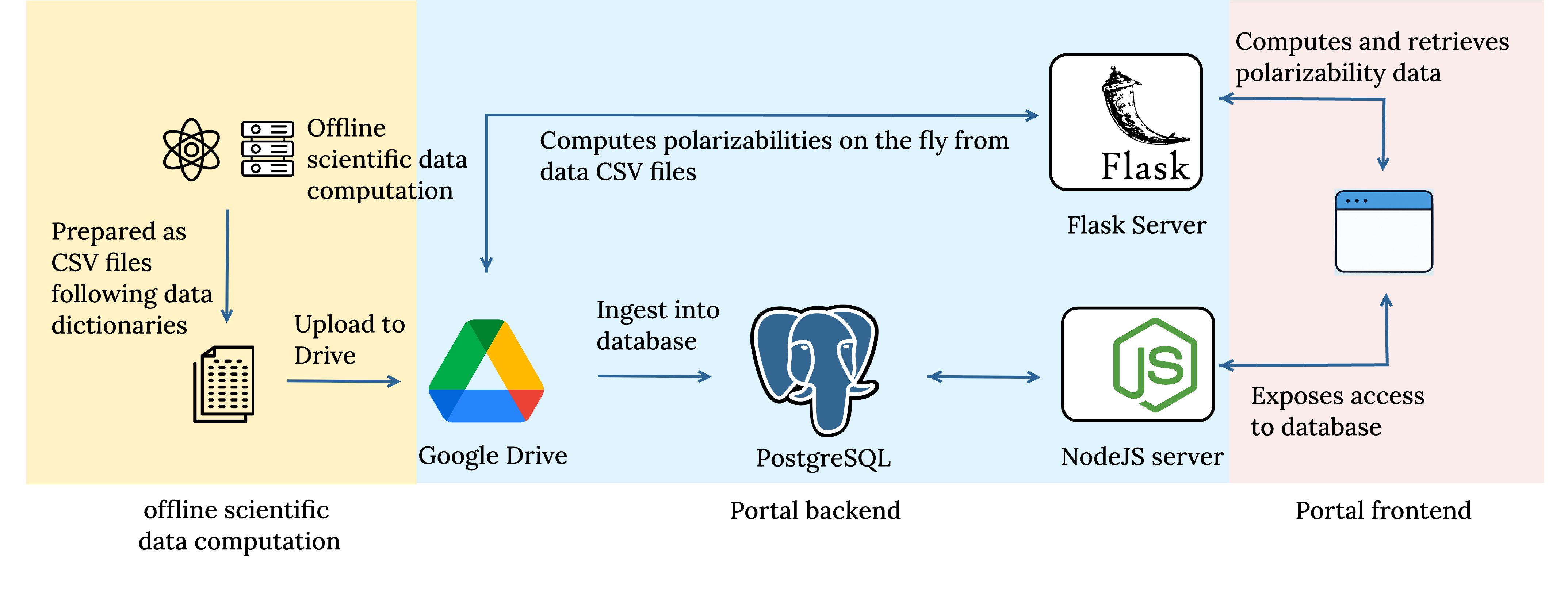}
\caption{\label{fig:arch}The overall architecture of the Atom portal. Portal data is generated from offline atomic computations, which are prepared as CSV files following pre-defined data dictionaries. The CSV files are uploaded to Google Drive, and then an automated database ingestion pipeline is triggered, which downloads the data to the portal backend, then automated scripts transform the data before ingesting them into a PostgreSQL database. A separate Flask service is designed for on-the-fly calculations of polarizabilities. The portal frontend allows users to access the database, as well as compute and retrieve polarizability data.}
\label{arch}
\end{figure*}

\section{Portal design}\label{sec:portal}
The portal is a full-stack web application designed to serve high-precision atomic data through a fast, interactive interface backed by an automated cloud-based data pipeline. It combines a modular React frontend with a Node.js backend and a separate Python Flask service. Data is stored in a PostgreSQL database and ingested through structured CSV uploads via Google Drive, allowing data contributions without direct interaction with the codebase. The portal design follows best practices~\cite{barakhshan2021exchanging} and lessons learned from related projects~\cite{GatewaySSI,amosgateway,MolSSI-project}. This section details the architecture, backend processing, dynamic computation of polarizabilities, and the testing infrastructure that ensures data validity across the system.

\subsection{High-level Architecture}
To realize the functionality introduced in Section~\ref{sec:UserInterface}, the portal uses a three-tier architecture, shown in Figure~\ref{fig:arch}: the end user interface (frontend), the server-side logic (backend), and the automated cloud data pipeline for ingesting new physics data into the portal. 

\paragraph{Cloud data pipeline}
To streamline data production workflows, the portal leverages Google Drive for data submission. This is important as the ease of use of Google Drive lowers the technical barrier to contributing data. Data is formatted according to pre-defined data dictionaries before uploading. These data dictionaries explicitly define the data schema and column types. Once the files are uploaded, they trigger an automated database ingestion pipeline for portal access. Once downloaded onto the backend, automated scripts then parse, transform, and validate the data according to scientific rules before ingesting them in a PostgreSQL database. 

\paragraph{Backend}
The primary backend is implemented using Node.js, providing RESTful APIs to manage data transactions, retrieval, and interactions with the PostgreSQL database. Additionally, a separate computational backend service implemented in Python using Flask is designed specifically for on-the-fly calculations of polarizabilities, magic wavelengths, and magic zeros -- this is because pre-computing and storing polarizability data is infeasible due to its large size. The backend is hosted on the University of Delaware's high-performance computing (HPC) resources, leveraging a customized server setup.

\paragraph{Frontend}
The frontend is designed with modular reusable components and complex state management using Next.js, a modern web framework. This is a significant improvement over the previous version of the portal, which was built with static HTML pages.

\subsection{Polarizabilities}
A separate backend using Flask is implemented to support polarizability calculations. Previous versions of the Atom Portal precomputed polarizabilities with CSV file sizes as large as 200,\,000 rows per element. This system became intractable once we added the feature of changing $\theta$ in the polarizability plots. We considered implementing this logic using the existing Node.js backend, but JavaScript has limited support for vector mathematics, which is necessary when performing algebraic reductions on such vast amount of data. Therefore, we opted for a separate Python Flask backend that could provide on-the-fly polarizability calculations directly from the CSV files, which are stored as blobs on the backend server. In addition to polarizability calculations, the Flask backend supports calculating tune-out (crossings with the $x$-axis) and magic wavelengths (crossings with polarizability graphs from different states), which are valuable for the scientific community. These are implemented using SciPy, a fast scientific data-processing Python library. 

Furthermore, to support extended graphing capabilities of polarizability data and a rich user interface, we modified Chart.js, a JavaScript graphing library. 

\subsection{Data testing}

The portal includes a testing framework to ensure consistency and correctness of the displayed data.

\paragraph{Internal data validation}
During the processing of the CSV files before database ingestion, processed data are compared against specially designated CSV files containing sample processed atomic data. Although this may seem redundant, because both data are derived from the same scientific rules, it is a highly valuable consistency check between the results of the atomic calculations and the Python code in the Portal backend. This test catches errors including precision and accuracy issues, uncertainty calculation mistakes, overlooked selection rules, 
and even incorrect values that were correctly calculated on the scientific Fortran code, but not by the portal Python code, or vice versa. 

\paragraph{Inter-version testing}
Because this new version of the portal contains data from previous versions, we also designed a testing mechanism implemented using web-scraping to test the consistency of the data between different versions. To accommodate the dynamic nature of the new version compared to the static pages of the older versions, we use Selenium, a headless browser that allows for scraping of dynamically generated webpages. On many occasions, these tests have surfaced issues that originate from the data files that were mishandled from version to version, or out-of-date data. 

\subsection{Data display}

\paragraph{References}
We also include a global key-value pair map for references to the data displayed on the portal. This map is stored in the database, where the keys are embedded in the CSV files. This system ensures accurate formatting and spelling of the references, which can fail if stored in raw text format.

\paragraph{Numerical representation}
One challenge is to ensure accurate numerical precision as the data are processed among multiple programming languages: Fortran, Python, SQL, and JavaScript. This challenge is overcome through the tests mentioned above and adherence to basic mathematical rules regarding precision under specific uncertainties. 

When uncertainties are available, we format the numbers in a compact display format: \verb|value(uncertainty)|.
For example, for transition probabilities, which are listed in scientific notation, 5.351(32)E+06 s$^{-1}$ means that the value 5.351E+06 s$^{-1}$ has an uncertainty of 0.032E+06 s$^{-1}$.

In cases where the uncertainties are larger than the corresponding value, we use the following display format: \verb|<value+uncertainty|. 
For example, 
an $E1$ matrix element with a value of 0.02397 a.u. and an uncertainty of 0.02756 a.u. is displayed as $<0.052$ a.u. 

Internally, we use a standardized approach to represent atomic states with ASCII, obviating the need for LaTeX representations and allowing representation in CSV tables. For example, ``\verb|5s2.5p, 2P, 1/2|'' encodes the electronic configuration, term symbol, and total angular momentum, respectively, corresponding to the LaTeX display: $5s^2 5p\,^2\!P_{1/2}$ .

\section{Conclusions and Future prospects}
In this paper, we introduce a free, open-access online portal for high-precision atomic data and computation (\href{https://www.udel.edu/atom}{https://www.udel.edu/atom}). The current portal provides atomic properties of 28 atoms and ions, including uncertainties where applicable for theoretical values. 
New workflow algorithms have been developed to facilitate generating large volumes of data for the portal, as well as to compare our calculated values with data from the NIST Atomic Spectra Database. We include an extensive testing framework to validate our calculated data, as well as ensure consistency and correctness of the displayed portal data.
The portal features an interactive polarizability plotting interface, as well as an automated data pipeline, allowing it to scale with new elements and data properties in future updates. 
Since the initial release, over 5,\,900 users have accessed and used the portal. 

We plan to continuously compute and add atomic data to the portal, with an initial goal of data for up to 100 atomic systems by Version 4.  Additional elements would include Be, Al$^+$,  Ba, Ra, Lu$^+$, In$^+$, Th$^{2+}$, Zn, Cd, Ti$^{+}$,  Ti, Sn$^{2+}$, Al$^{2+}$, Sc$^{2+}$, Y$^{2+}$, Lu$^{2+}$, La$^{-}$, La$^{2+}$, Th$^{3+}$, Fe$^{16+}$, In, Tl, and many others, including highly charged ions. We also plan to continue to collect and display precision experimental data, including data for Yb$^+$ and Yb. 
Furthermore, the capabilities of the portal can be expanded to include new features, including different data representations and search options.

\section{Acknowledgments}
\label{sec:acknowledgments}
We thank Adam Marrs for work on Version 1 of the portal. 
We thank Sergey Porsev, Andrey Bondarev, and many others for thoroughly testing the database.
The developments and calculations in this work were done through the use of IT resources at the University of Delaware, specifically the high-performance Caviness and DARWIN~\cite{DARWIN} computer clusters. This work was supported in part by the National Science Foundation under Awards No. OAC-1931339 and OAC-2209639. VB is grateful for the hospitality of Perimeter Institute where part of this work was carried out. Research at Perimeter Institute is supported in part by the Government of Canada through the Department of Innovation, Science and Economic Development and by the Province of Ontario through the Ministry of Colleges and Universities.


\bibliographystyle{elsarticle-num}
\bibliography{refs,re}

\newpage

\appendix

\section{pCI automated scripts\label{pCI-py}}
The pCI code package~\cite{2025pCI} contains a library of Python scripts that automate parts of the methods mentioned above. These scripts were used to generate volumes of atomic data for the portal. The sequence of scripts (\verb|basis.py|, \verb|ci.py|, \verb|dtm.py|) 
performs the CI, CI+MBPT, or CI+all-order methods, generating all relevant data for the atomic system of interest. We leave details of these scripts to Ref.~\cite{2025pCI}.
In addition, we developed new workflow scripts specific to the portal: \verb|gen_portal_csv.py| and \verb|calc_lifetimes.py|. These scripts can also be found in the \verb|py-lib| directory of the pCI distribution~\cite{2025pCI}. In the following, we discuss some technical details of these scripts.

\subsection{gen\_portal\_csv.py}
The \verb|gen_portal_csv.py| script generates CSV files of atomic energy levels and matrix elements given output files from the \verb|pconf| and \verb|pdtm| programs. This script reads these output files, containing energies and transition matrix elements, compares the results with the NIST database, then compiles all relevant atomic data in a format readable by the atomic portal. This script reads the portal block from \verb|config.yml|:
\begin{verbatim}
    portal:
        ignore_g: True
        min_uncertainty: 1.5 
\end{verbatim}

The \verb|portal.ignore_g| field ignores all computed atomic properties with $g$ in the configuration or $G$ in atomic term symbols.
The \verb|portal.min_uncertainty| field sets a minimum uncertainty in percentage for matrix value uncertainties. Pre-defined minimum uncertainties are already set for a few systems, estimating percentage contribution from higher-order computations. If this value is set, the matrix element uncertainty is replaced with a new value calculated in quadrature with the the minimum uncertainty specified. 

The method of generating atomic data files for the portal is separated into three major steps: 

\textit{1. Reading and reformatting input files. } 
At the first stage, all computed energy levels are read from the \verb|FINAL.RES| output files of the \verb|pconf| program. The script will first check generated directories from the \verb|ci.py| script for these files. If they exist, the script will place them in the \verb|DATA_RAW| directory.
If they are not detected, it is up to the user to place them in the \verb|DATA_RAW| directory. The script looks for results from \verb|FINALeven.RES| and \verb|FINALodd.RES| for CI+all-order calculations, and \verb|FINALevenMBPT.RES| and \verb|FINALoddMBPT.RES| for CI+MBPT calculations. The ``even'' and ``odd'' designation refers to the parity of the configurations. 
In addition, electric-dipole matrix element output files \verb|E1.RES| can be read. In this case, the same checks are made, but now in the generated directories from the \verb|dtm.py| script. The CI+all-order and CI+MBPT output files should be named \verb|E1.RES| and \verb|E1MBPT.RES|. 

Once all input files are read, then the \verb|gen_portal_csv.py| script will check if both CI+all-order and CI+MBPT results exist. If they do exist, then uncertainties are calculated as the difference between the two results for each energy and matrix element. At this point, a new CSV file \verb|FINAL.csv| is also written with uncertainties attached. This is essentially \verb|FINAL.RES| with uncertainties. Otherwise, if only one set of calculations were found, uncertainties are set to 0. 

\textit{2. Filtering and correcting misidentified configurations. }
Sometimes, the configurations and terms of energy levels outputted by the pCI codes might be misidentified due to discrepancies in the basis set or numerical precision. At this stage, we attempt to correct misidentified configurations using the data acquired from the NIST database. Here, we create a correspondence that maps the parsed NIST identifications to the pCI computed data. This correspondence, or mapping, is written to the file \verb|Element_parity.txt| in the \verb|DATA_Output| directory, for each parity. We describe the correspondence with NIST algorithm in more detail in Sec.~\ref{sec:correspondence}.

\textit{3. Outputting data for portal. }
At final stage of the \verb|gen_portal_csv.py| script, the mapping of NIST and pCI energy levels is reformatted for use on the portal. The output file is a \verb|Element_Energies.csv|, a CSV file of the energies of the system of interest, with a preference for NIST data over computed data, i.e. for each configuration, if NIST energy and identification exists, theory data is replaced by it. The final column of this file, \verb|is_from_theory|, is set to \verb|True| if theory values were kept, and no mapping was made to available NIST data. Matrix elements are stored in \verb|Element_Matrix_Elements_Theory.csv|, and transition rates are stored \verb|Element_Transition_Rates.csv|. Note that at the step when transition rates are calculated, NIST identifications and energies have replaced pCI computed data, so final transition rates are calculated using NIST data, when available. Additionally, the \verb|calc_lifetimes.py| script can be run to generate CSV files of lifetimes and transition rates given output files from \verb|gen_portal_csv.py|. These are outputted to the files \verb|Element_Lifetimes_Error_Check.csv| and \verb|Element_Transition_Rates_Error_Check.csv|, respectively. The \verb|*_Error_Check.csv| files are formatted by the data dictionaries discussed in Sec.~\ref{sec:portal}.







\end{document}